\documentclass[sigconf]{acmart}

\AtBeginDocument{%
  \providecommand\BibTeX{{%
    \normalfont B\kern-0.5em{\scshape i\kern-0.25em b}\kern-0.8em\TeX}}}

\usepackage{amsfonts}
\usepackage{graphicx}
\usepackage{subfigure}
\usepackage{multirow}
\usepackage{algorithm}
\usepackage{algorithmic}
\usepackage{enumitem}
\usepackage{float}
\usepackage{hyperref}
\usepackage{xspace}
\usepackage[title]{appendix}

\copyrightyear{2023}
\acmYear{2023}
\setcopyright{acmlicensed}\acmConference[WWW '23]{Proceedings of the ACM Web Conference 2023}{May 1--5, 2023}{Austin, TX, USA} \acmBooktitle{Proceedings of the ACM Web Conference 2023 (WWW '23), May 1--5, 2023, Austin, TX, USA}
\acmPrice{15.00}
\acmDOI{10.1145/3543507.3583467} \acmISBN{978-1-4503-9416-1/23/04}

\begin{document}

\title{Multi-Task Recommendations with Reinforcement Learning}

\author{Ziru Liu}
\affiliation{%
\institution{City University of Hong Kong}
\country{}}
\email{ziruliu2-c@my.cityu.edu.hk}

\author{Jiejie Tian}
\affiliation{%
\institution{City University of Hong Kong}
\country{}}
\email{jiejitian2-c@my.cityu.edu.hk}

\author{Qingpeng Cai*}
\affiliation{%
\institution{Kuaishou Technology}
\country{}}
\email{cqpcurry@gmail.com}

\author{Xiangyu Zhao*}
\affiliation{%
\institution{City University of Hong Kong}
\country{}}
\email{xianzhao@cityu.edu.hk}

\author{Jingtong Gao}
\affiliation{%
\institution{City University of Hong Kong}
\country{}}
\email{jt.g@my.cityu.edu.hk}

\author{Shuchang Liu}
\affiliation{%
\institution{Kuaishou Technology}
\country{}}
\email{liushuchang@kuaishou.com}

\author{Dayou Chen}
\affiliation{%
\institution{City University of Hong Kong}
\country{}}
\email{dayouchen2-c@my.cityu.edu.hk}

\author{Tonghao He}
\affiliation{%
\institution{City University of Hong Kong}
\country{}}
\email{tonghaohe2-c@my.cityu.edu.hk}

\author{Dong Zheng}
\affiliation{%
\institution{Kuaishou Technology}
\country{}}
\email{zhengd07@qq.com}

\author{Peng Jiang}
\affiliation{%
\institution{Kuaishou Technology}
\country{}}
\email{jp2006@139.com}

\author{Kun Gai}
\affiliation{%
\institution{Unaffiliated}
\country{}}
\email{gai.kun@qq.com}

\thanks{* Corresponding authors.}



\renewcommand{\shortauthors}{Liu, et al.}

\begin{abstract}
In recent years, Multi-task Learning (MTL) has yielded immense success in Recommender System (RS) applications \cite{silver2014deterministic}.
However, current MTL-based recommendation models tend to disregard the session-wise patterns of user-item interactions because they are predominantly constructed based on item-wise datasets.
Moreover, balancing multiple objectives has always been a challenge in this field, which is typically avoided via
linear estimations in existing works.
To address these issues, in this paper, we propose a Reinforcement Learning (RL) enhanced MTL framework, namely RMTL, to combine the losses of different recommendation tasks using dynamic weights.
To be specific, the RMTL structure can address the two aforementioned issues by (\textit{i}) constructing an MTL environment from session-wise interactions and (\textit{ii}) training multi-task actor-critic network structure, which is compatible with most existing MTL-based recommendation models, 
and (\textit{iii}) optimizing and fine-tuning the MTL loss function using the weights generated by critic networks. Experiments on two real-world public datasets demonstrate the effectiveness of RMTL with a higher AUC against state-of-the-art MTL-based recommendation models.
Additionally, we evaluate and validate RMTL's compatibility and transferability across various MTL models. 

\end{abstract}



\maketitle

\vspace{-1mm}
\section{Introduction}
The evolution of the Internet industry has led to a tremendous increase in the information volume of online services ~\cite{aceto2020industry}, such as social media and online shopping. In this scenario, the Recommender System (RS), which distributes various types of items to match users' interests, has made significant contributions to the enhancement of user online experiences in a variety of application fields, such as products recommendation in e-commerce platforms, short video recommendation in social media applications ~\cite{ma2018esmm,tang2020ple}. 
In recent years, researchers have proposed numerous techniques for recommendations, including collaborative filtering~\cite{mooney2000content}, matrix factorization based approaches~\cite{koren2009matrix}, deep learning powered recommendations~\cite{cheng2016wide,zhang2019deep}, etc. The primary objective of RS is to optimize a specific recommendation object, 
such as click-through rate and user conversion rate.
However, users usually have varying interaction behaviors on a single item. 
In short-video recommendation services, for instance, users exhibit a wide range of behavior indicators, such as clicks, thumbs, and continuous dwelling time ~\cite{huang2016real};  
while in e-commerce platforms, the developers not only focus on the users' clicks but also on the final purchases to guarantee profits. 
All these potential issues prompted the development of Multi-Task Learning (MTL) techniques for recommender systems in research and industry communities ~\cite{vithayathil2020survey,ma2018mmoe,tang2020ple}.

MTL-based recommendation models learn multiple recommendation tasks simultaneously by training in a shared representation and transferring information among tasks \cite{caruana1997multitask}, which has been developed for a wide range of machine learning applications, including computer vision ~\cite{ren2015faster}, natural language processing ~\cite{goldberg2017neural}, click-through rate (CTR) and click-through\&conversion rate (CTCVR) prediction ~\cite{ma2018mmoe}. 
The objective functions for most existing MTL works are typically linear scalarizations of the multiple-task loss functions ~\cite{ma2018esmm,ma2018mmoe,tang2020ple}, which fix the weight with a constant. This item-wise multi-objective loss function is incapable of ensuring the convergence of the global optimum and typically yields limited prediction performance. 
On the other hand, at the representation level, the input of most existing MTL models is assumed to be the feature embeddings and user-item interaction (called item-wise), despite the fact that sequentially organized data (i.e., session-wise inputs) are relatively more prevalent in real-world RS applications. For example, the click and conversion behaviors of short video users typically occur during a specific session, so their inputs are also timing-related.
However, this will downgrade the MTL model performance, while some tasks may have conflicts between session-wise and item-wise labels ~\cite{caruana1997multitask}. Exiting MTL models concentrate on the design of network structures to improve the generalization ability of the model, while the study of proposing a new method that enhances the multi-task prediction weights considering the session-wise patterns has not received sufficient attention.

To address the two above-mentioned problems, we propose an RL-enhanced multi-task recommendation framework, RMTL, which is capable of incorporating the sequential property of user-item interactions into MTL recommendations and automatically updating the task-wise weights in the overall loss function. 
Reinforcement Learning (RL) algorithms have recently been applied in the RS research, which models the sequential user behaviors as Markov Decision Process (MDP) and utilizes RL to generate recommendations at each decision step ~\cite{zhao2011reinforcement, mahmood2007learning}. The RL-based recommender system is capable of handling the sequential user-item interaction and optimizing long-term user engagement~\cite{afsar2021reinforcement}.
Therefore, our RL-enhanced framework RMTL can convert the session-wise RS data into MDP manner, and train an actor-critic framework to generate dynamic weights for optimizing the MTL loss function. 
To achieve multi-task output, we employ a two-tower MTL backbone model as the actor network, which is optimized by two distinct critic networks for each task. 
In contrast to existing MTL models with item-wise input and constant loss function weight design, our RMTL model extracts sequential patterns from session-wise MDP input and updates the loss function weights automatically for each batch of data instances. 
In this paper, we focus on the CTR/CTCVR prediction, which is a crucial metric in e-commerce and short video platform ~\cite{liu2020kalman}. Experiments against state-of-the-art MTL-based recommendation models on two real-world datasets demonstrate the effectiveness of the proposed model.

We summarize the contributions of our work as follows:
(i) The multi-task recommendation problem is converted into an actor-critic reinforcement learning scheme, which is capable of achieving session-wise multi-task prediction;
(ii) We propose an RL-enhanced Multi-task learning framework RMTL, which can generate adaptively adjusted weights for loss function design. RMTL is compatible with most existing MTL-based recommendation models; 
(iii) Extensive experiments on two real-world datasets demonstrate the superior performance of RMTL than SOTA MTL models, we also verify RMTL's transferability across various MTL models.


\begin{figure*}
    \Description{Overview of the RMTL framework. Including the actor-critic network structures.}
    \centering
    \includegraphics[width=0.78\linewidth]{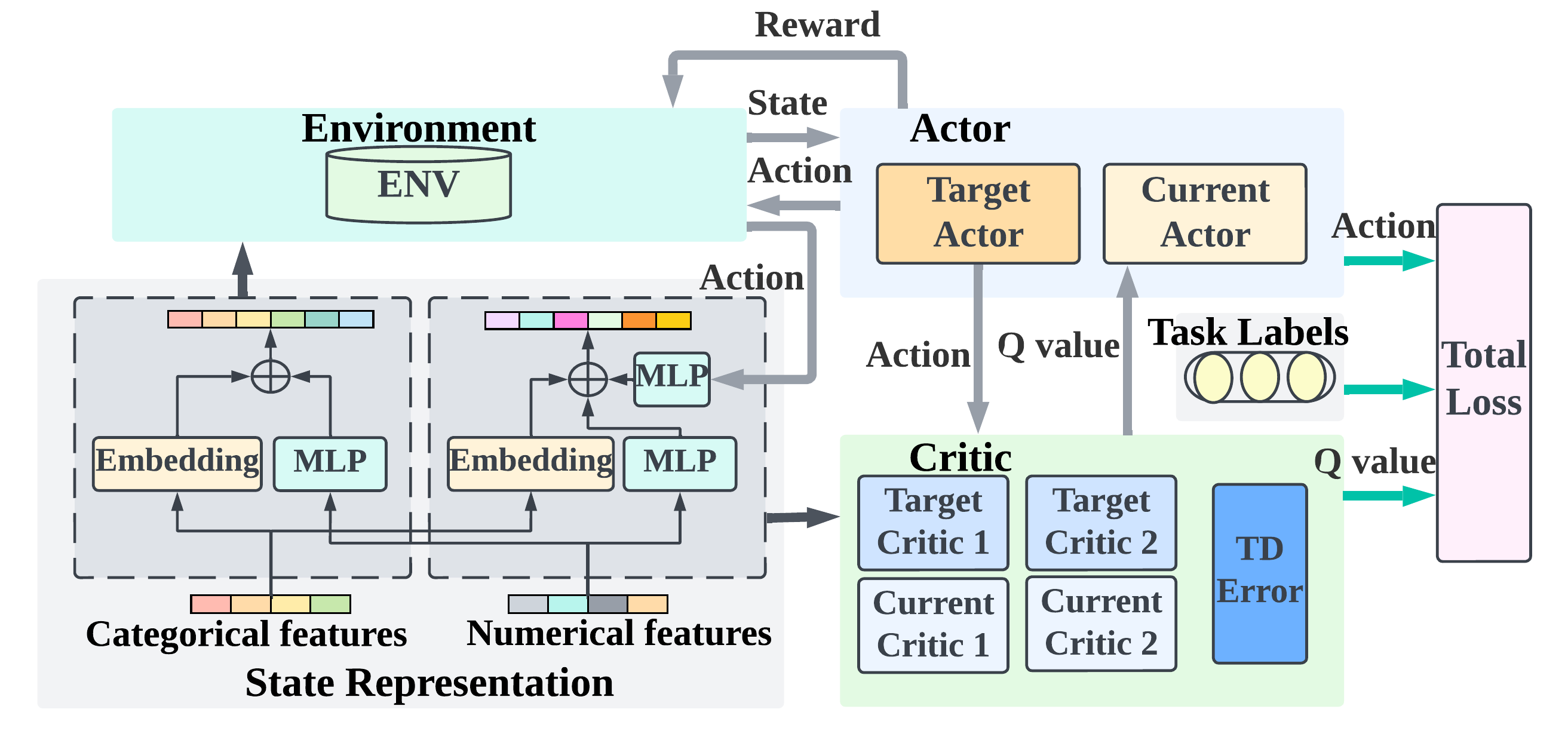}
    \vspace*{-5mm}
    \caption{Overview of the RMTL framework.}
    \label{fig:Frame}
    \vspace*{-5mm}
\end{figure*}
\section{THE PROPOSED Framework}
This section will give a detailed description of our method, the RMTL framework, which effectively addresses the bottlenecks of exiting works by realizing session-wise multi-task prediction with dynamically adjusted loss function weights.

\subsection{Preliminary and Notations}

{\textbf{Session-wise Multi-task Recommendations.}} We note that the choice of loss function as an item-wise multi-objective combination may lack the ability to extract sequential patterns from data. In our work, we propose the session-wise multi-objective loss, which optimizes the objective by minimizing weighted cumulative loss at each session. Given a $K$-task\footnote{Note that for multi-task recommendations in CTR/CTCVR prediction setting, we have $K=2$.} prediction dataset 
with $D:= \left\{user_n, item_n, (y_{n,1},...,y_{n,K})\right\}_{n=1}^N $, which consist of $N$ user-item interaction records, where $user_n$ and $item_n$ means the $n$-th user-item $id$. Each data record has $K$ binary 0-1 labels ${y_1,\cdots,y_K}$ for the corresponding task. We define the session as $\left\{\boldsymbol{\tau}_m \right\}_{m=1}^M$, each session $\boldsymbol{\tau}_m$ contains $T_m$ different discrete timestamps: $\tau_m = \left\{ \tau_{m,1},\cdots,\tau_{m,t},\cdots,\tau_{m,T_m} \right\} $, where $\tau_{m,t}$ stands for the 
timestamp $t$ in session $m$. The corresponding label is denoted by $\left\{ (y^s_{t,1},...,y^s_{t,K})\right\}$.
The session-wise loss function for all sessions parameterized by $\theta_1^s,...,\theta_K^s$ is defined as:
\begin{equation} \label{eq:swl}
    \begin{aligned}
        \underset{\left\{ \theta_1^s,...,\theta_K^s \right\}} {\arg \min} \mathcal{L}^s(\theta_1^s,\cdots, \theta_K^s) = \sum\limits_{k=1}^K  ( \sum\limits_{m=1}^M \sum\limits_{t=1}^{T_m} \omega^s_{k, \tau_{m,t}} L^s_{k, \tau_{m,t}} (\theta_k^s) ) 
        \\ s.t. \quad L^s_{k,\tau_{m,t}}(\theta_k^s)= - [y^s_{k,\tau_{m,t}} log(z^s_{k,\tau_{m,t}}(\theta^s_k)) + 
        \\ (1-y^s_{k,\tau_{m,t}}) log(1-z^s_{k,\tau_{m,t}}(\theta_k^s))]
    \end{aligned}
\end{equation}
where $\mathcal{L}^s(\theta_1^s,\cdots, \theta_K^s)$ is the session-wise total loss function, and $\omega^s_{k, \tau_m}$ is the corresponding weight for the BCE loss, which adaptively adjusted by the system. $z^s_{n,k}(\theta_k^s)$ is the session-wise prediction value at $n$-th data point for task $k$ parameterized by $\theta_k^s$. In order to obtain such weight, we take advantage of the reinforcement learning framework to estimate the weight using several dynamic critic networks that could tune their outputs at different solving steps of the objective function.

\subsection{Framework Overview}
We present the framework of RMTL in Figure ~\ref{fig:Frame} with a state representation network, an actor network, and critic networks. The learning process of RMTL can be summarized as two steps:
\begin{itemize}[leftmargin=*]
    \item \textbf{The Forward Step.} Given the user-item combined features, the state representation network generates the state $s_t$ based on the input feature at timestamp $t$. Then, we get the action $(a_{1,t},a_{2,t})$ from the actor network extracting state information. The action value $(a_{1,t},a_{2,t})$ and user-item combined features are further processed by the MLP layer and embedding layer as the input of the critic network for calculating Q-value from critic network $Q(s_t, a_{k,t}; \mathbf{\phi}_k)$ for each task $k$. Finally, the  overall loss function for multi-task $\mathcal{L}(\theta_1,\theta_2)$ can be estimated according to BCE loss for each task and adapted weight $\omega_{k,t}$ controlled by $Q(s_t,a_{k,t};\phi_k )$.
    \item \textbf{The Backward Step.} We first update the critic network parameters $\phi_k$ based on TD error $\sigma$ and gradients of Q-value. Then, we optimize the parameters of the actor network $\theta_k$ with respect to the overall loss function. 
\end{itemize}

\subsection{Markov Decision Process (MDP) Design}
\label{2.3}
We translate the multi-task prediction data into an MDP format $(\mathcal{S}, \mathcal{A}, \mathcal{P}, \mathcal{R}, \gamma)$ to analyze in the reinforcement learning framework.
\begin{itemize}[leftmargin=*]
    \item \textbf{State space $\mathcal{S}$} is the set of state $s_t \in \mathbf{R}^d$ containing the user-item combined features, where $d$ is the combined feature dimension; 
    
    \item \textbf{Action space} In our CTR/CTCVR prediction setting, $\mathcal{A}$ is the set of continuous action pairs $(a_{1,t},a_{2,t}) \in [0,1]^2$ containing the actions for two specific prediction tasks, where each element in $\mathcal{A}$ represents prediction value for CTR and CTCVR; 
    
    \item \textbf{Reward} $r_{k,t}=r(a_{k,t},y_{k,t}) \in \mathbf{R}, k = 1,2$, is the feedback from the environment related to the current actions and the ground-truth click/pay label pair $(y_{1,t},y_{2,t})$ of the current item. In order to be consistent with the definition of BCE loss, we define the reward function for each step using the negative BCE value, i.e.,
    \begin{equation} \label{eq:bce-reward}
        r_{k,t} = y_{k,t} log(a_{k,t}) + (1-y_{k,t}) log(1-a_{k,t}), \quad k = 1,2 
    \end{equation}
    \item \textbf{Transition} $\mathcal{P}_{sas'} = \chi(s' = s_{t+1})$ is the transition probability from state $s$ to state $s'$ based on the user interaction sequence, and $\chi$ is the indicator function that set the probability as $1$ when the next state (item) corresponds to the sequence;
    \item \textbf{Discount} $\gamma$ the discount rate is set as 0.95 for our case.
\end{itemize}

The RMTL model learns a optimal policy $\pi(s_t; \mathbf{\theta}^*)$ which converts the item-user feature $s_t \in \mathcal{S}$ into the continuous action value $(a_{1,t},a_{2,t}) \in \mathcal{A}$ to maximize the weighted total rewards:
\begin{equation} \label{eq:tr}
    \begin{aligned}
        \underset{\left\{ \theta_1, \theta_2 \right\}}{\max} \sum_{k=1}^2 
        (\sum\limits_{m=1}^M \sum\limits_{t=1}^{T_m} \omega^r_{k, \tau_{m,t}} R_{k, \tau_{m,t}} (\theta_k))
        \\ s.t. \quad R_{k, \tau_{m,t}}(\theta_k)= r(a_{k, \tau_{m,t}},y_{k, \tau_{m,t}}) \quad k=1,2
    \end{aligned}
\end{equation}
where $\omega^r_{k, \tau_{m,t}}$ is the weight for reward at session $\tau_m$ in timestamp $t$ for task $k$. Since the reward function defined by Equation (\ref{eq:bce-reward}) is the negative BCE, and action value is generated from policy parameterized by $\theta_k$. The optimization problem above is equivalent to minimizing the session-wise loss function in Equation (\ref{eq:swl}).

\subsection{Session MDP Construction} 
Based on the MDP design in Section \ref{2.3}, we construct session MDP for the RL training, which can improve the performance of the MTL model. 
Classic MTL methods usually face challenges in introducing sequential user behaviors into the modeling, where the timestamps of user behaviors are highly correlated. Reinforcement learning built upon the MDP sequences can be a solution to address this problem. However, the alignment order of MDP may have great influences on the performance of RL, since the decision-making of the next action depends on the temporal difference (TD). In this paper, we construct the session MDP, which is organized by \textit{session id}. For each session $\tau_m$, the transition records are separated by the timestamps stored in the original dataset. This construction generates session MDP sequences organized by sequential user behaviors and has the advantage of overall loss weight updating.


For CTR/CTCVR prediction task, i.e., $K=2$, assume we have a session-wise dataset $D_s$ $= \{S_{\tau_m}\mathbf{1}_{\tau_m},  U_{\tau_m}\mathbf{1}_{\tau_m}, \textbf{I}_{\tau_m}, (\textbf{y}_{1,\tau_m}, \textbf{y}_{2,\tau_m}) \}_{m=1}^M$
consisting of $M$ session records, where $S_{\tau_m}$ and $U_{\tau_m}$ are the $m$-th session and user $id$. $\textbf{I}_{\tau_m}$ is item $id$ vector whose dimension equals the item number that interacted with the user in $m$-th session. Each record stores the user-item information $\textbf{x}_{\tau_m} = \{U_{\tau_m} \mathbf{1}_{\tau_m}, \textbf{I}_{\tau_m}\}$ \footnote{ $\mathbf{1}_{\tau_m}$ is the unit vector with dimension equals the item number in $m$-th session. } 
and corresponding binary 0-1 label vector $(\textbf{y}_{1,\tau_m}, \textbf{y}_{2,\tau_m})$ for click and convert indicators. Use feature vector generated from mapping function $f$ for further processing by state function$F$, then input to agent policy $\pi(s; \theta_k)$, which is a pretrained MTL network. The replay buffer $\mathcal{B}$ contains $M$ session MDP sequences $\{(s_{\tau_m,1},\cdots,s_{\tau_m,T_m}) \}_{m=1}^M$, each of them have $T_m$ transition information based on session $\tau$. The detail of session MDP construction is shown in Algorithm \ref{alg:env-sess}. Specifically, we separate each session-wise data based on timestamps. Then, randomly sample a session and generate state information by the functions $f$ and $F$ (line 2). The state information of records in a specific session are sequentially input into the agent policy $\pi(s; \theta_k)$, which generates estimated action pairs $(a_{1,t},a_{2,t})$ (line 4). Then, we further calculate the reward value according to the Equation (\ref{eq:bce-reward}) and store the transition $(s_t,a_{1,t},a_{2,t},s_{t+1},r_{1,t},r_{2,t})$ at timestamp $t$ into the experience replay buffer $\mathcal{B}$, which provides a more stable training environment (line 5). This completes the session-wise CTR/CTCVR prediction setup for the RL training.

\subsubsection{\textbf{State Representation Network.}} \label{sss:srn}
The original features of item-user pairs are categorical (represented by one-hot encoding format) or numerical. The state representation network is the combination of embedding layer and multi-layer perceptron to extract the user-item features, which corresponds to the mapping function $F$ in Algorithm \ref{alg:env-sess}. The categorical features are firstly converted into binary vectors with length $l_k$, and then input to the embedding layer $R^{l_k} \rightarrow R^{d_e}$ with embedding dimension $d_e$. Besides, the numerical features are converted to the same dimension with $d_e$ by a linear transformation. The features translated by the above process are combined and further used as input for the MLP network, which is the fully-connected neural network with multiple layers: 
\begin{equation} \label{eq:mlp} 
    \begin{aligned}
        \textbf{h}_{l+1} = \sigma(\textbf{W}_l\textbf{h}_l + \textbf{b}_l) \quad n=0,1,...,N-1  \\
        \textbf{h}_{L} = \sigma^*(\textbf{W}_{L-1}\textbf{h}_{L-1} + \textbf{b}_{L-1}) \qquad \qquad
    \end{aligned}
\end{equation}
where $\textbf{h}_l$ is the $l$-th hidden layer with weight $\textbf{W}_l$ and bias $\textbf{b}_l$, the corresponding activation function $\sigma$ is ReLU. The output layer is denoted by $\textbf{h}_L$ with the sigmoid function $\sigma^*$. The output $\textbf{h}_L$ is embedded with the extracted feature and represented by $F(f(x_t))$.

\begin{algorithm}[t]
\caption{Session-wise CTR/CTCVR Prediction Setup}
\label{alg:env-sess}
\raggedright
{\bf Input}: Session-wise dataset organized as $\{S_{\tau_1}:(\textbf{x}_{\tau_1},\textbf{y}_{1,\tau_1}, \textbf{y}_{2,\tau_1}), \cdots, S_{\tau_M}:(\textbf{x}_{\tau_M},\textbf{y}_{1,\tau_M}, \textbf{y}_{2,\tau_M})\}$ \\
{\bf Reorganize}: Separate each session with state-label pairs denoted as $(f(x_{\tau_{m,t}}),y_{1,\tau_{m,t}},y_{2,\tau_{m,t}})$, where $f$ is a mapping function that returns the feature vector for any given user-item $id$\\
\begin{algorithmic}[1]
\WHILE{Environment not closed}
\STATE \textbf{Reset}: randomly sample a session $\boldsymbol{\tau}_m$ and initialize state $s_1$ with $F(f(x_1))$, where $F$ is state representation function.
\WHILE{Session not done}
    \STATE \textbf{Input:} Use action value $a_{1,t},a_{2,t}$ from agent policy $\pi(s; \mathbf{\theta}_k)$ as prediction values for CTR/CTCVR
    \STATE \textbf{Step:} Compute the reward $r_{1,t}, r_{2,t}$ according to \ref{eq:bce-reward} with actions and labels, return next state $s_{t+1}$. Store the transition $(s_t,a_{1,t},a_{2,t},s_{t+1},r_{1,t},r_{2,t})$ into replay buffer $\mathcal{B}$
    \IF{$t+1 > T_m$ }
    \STATE Finish the session $\boldsymbol{\tau}_m$
    \ENDIF
\ENDWHILE
\ENDWHILE
\end{algorithmic}
\end{algorithm}

\subsection{RMTL Framework}
The RMTL aims to optimize the overall loss function with adaptive dynamic weights.
In this direction, existing works \cite{tang2020ple} try to design the constant rate updating strategy for the loss of weight by:
\begin{equation}
    \omega'_{k,t} = \omega'_{k,0} \gamma'^{t}
\end{equation}
where $\omega'_{k,t}$ is the weight for task $k$ at timestamp $t$, and $\gamma'$ is the constant decay rate. However, this weight is only controlled by time $t$ and has no interaction with the type of task. To address this problem, we apply the critic network for evaluating the estimated total reward $V_k$ of action $a_k$ generated from agent policy for each task. Then, the value of loss function weight for each task is calculated as the linear transformation with polish variable $\lambda$. This novel overall loss design can optimize the MTL model parameter in a better direction, enhancing the efficiency and prediction performance.

The reinforcement learning-based CTR/CTCVR prediction training process consists of two major ingredients: actor network and task-specific critic network. The actor network is the main structure for predicting CTR/CTCVR in our work, which can be thought of as the agent policy $\pi(s_t; \mathbf{\theta}_k)$ parameterized by $\theta_k$. Based on this actor network, the agent can choose an action for each state. The critic network denoted by $Q(s_t, a_{k,t}; \mathbf{\phi}_k)$ 
is the differentiable action value network updated by the gradient of Q-value and TD error $\delta$, which can observe the potential rewards of the current state by learning the relationship between the environment and the rewards. Besides, it generates the adaptively adjusted weights of BCE loss for the objective function and updates the actor network parameter $\theta_k$ along the direction of better policy decisions. 

\begin{figure}[t]
    \Description{Structure of Actor Network. The main modules are several MLPs.}
    \centering
    \includegraphics[width=1.\linewidth]{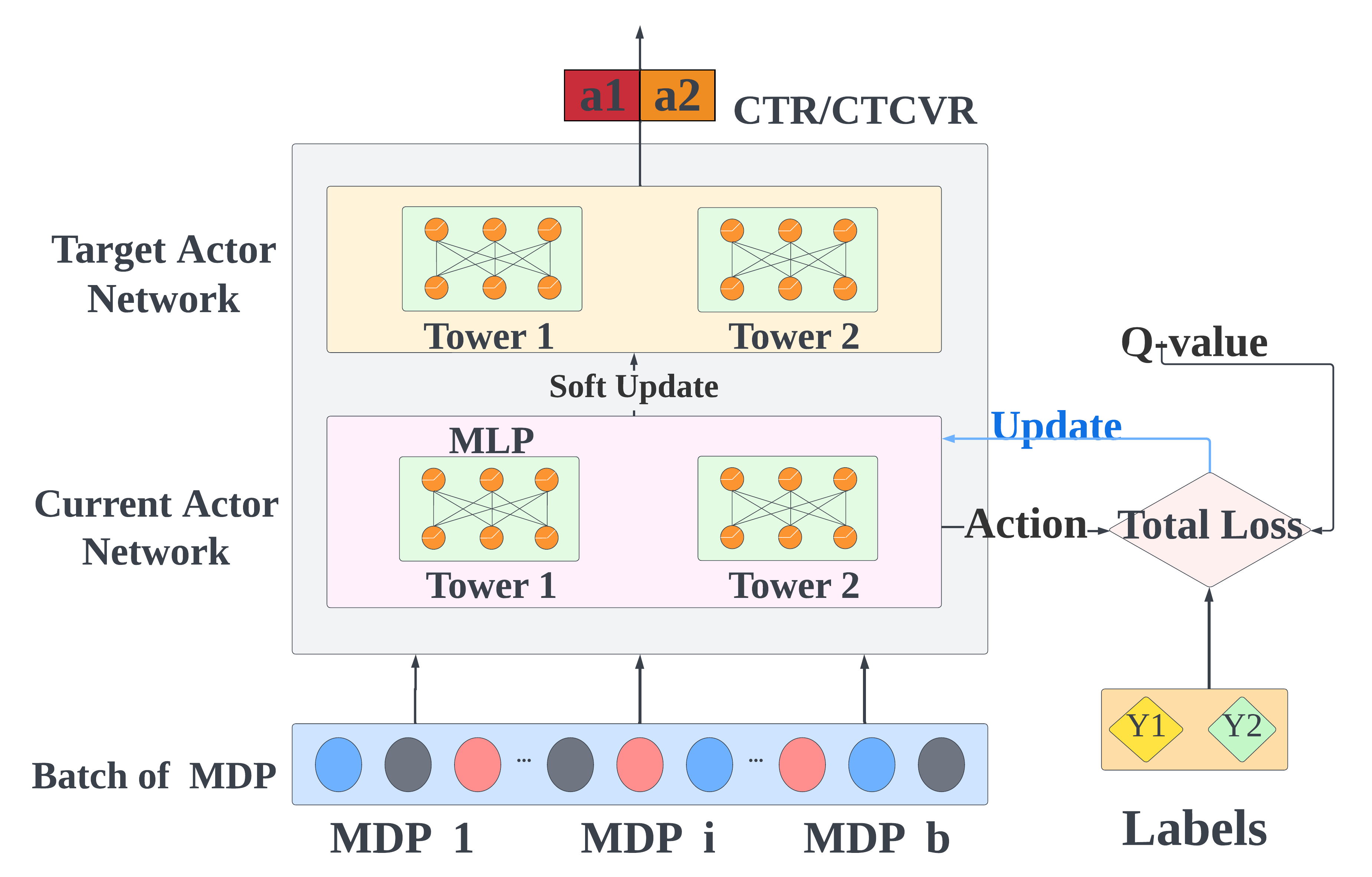}
    \vspace{-8mm}
    \caption{Structure of Actor Network. }
    \label{fig:Actor}
    \vspace{-6mm}
\end{figure}


\subsubsection{\textbf{Actor Network.}}
The actor network in the RL setting can be referred to as a policy that is represented by a probability distribution $\pi(s; \mathbf{\theta}_k) = \mathbf{P}(a|s,\theta_k)$b. In practice, this method usually suffers from two problems: (i) The actor critic network performs poorly in a setting with a large action space, and (ii) the estimated agent policy is trained by its own expected action value that is not the ground truth data. In this section, we address the above two problems by applying a specific MTL model as the agent policy with deterministic action output, which reduces the capacity of action space. In addition, we update the neural network based network parameters using ground truth task labels. 
The structure of actor network is shown in Figure \ref{fig:Actor}, which has a similar structure to specific MTL models, we may specify it using the ESMM model in this subsection.
The shared-bottom layer in the original actor network design is removed and set as a state representation network mentioned in Sub-subsection \ref{sss:srn} for generating replay buffer $\mathcal{B}$. Given a batch of MDP sequences from the replay buffer $\mathcal{B}$, we input the state information $s_t$ into the actor network, which is two parallel neural networks denoted by two tower layers parameterized by $\theta_1$ and $\theta_2$,
\begin{equation}
    \pi(s_t;\theta_k) = a_{k,t}, \quad \quad k=1,2
\end{equation}
The output of each tower layer is the deterministic action value, which represents the prediction for the specific task. In our setting, one tower outputs the CTR prediction value, and the other outputs the CTCVR prediction value. After the training process for the batch of MDP sequences, we calculate the  overall loss function based on the weighted summation of BCE loss:
\begin{equation}
    \mathcal{L}(\theta_1,\theta_2) = \sum\limits_{(s_t,\cdots,s_{t+1},\cdots)\in b} \sum\limits_{k=1}^2 \omega_{k,t} BCE(\pi(s_t; \mathbf{\theta}_k),y_{k,t}))
\end{equation}
This differentiable  overall loss function $\mathcal{L}(\theta_1,\theta_2)$ is controlled by binary cross entropy between estimated action and real task labels $y_{k,t}$, which enhances the decision accuracy of the policy.

\subsubsection{\textbf{Critic Network.}}
\begin{figure}[t]
    \Description{Structure of Critic Network. The main modules are several MLPs.}
    \flushleft
    \includegraphics[width=1\linewidth]{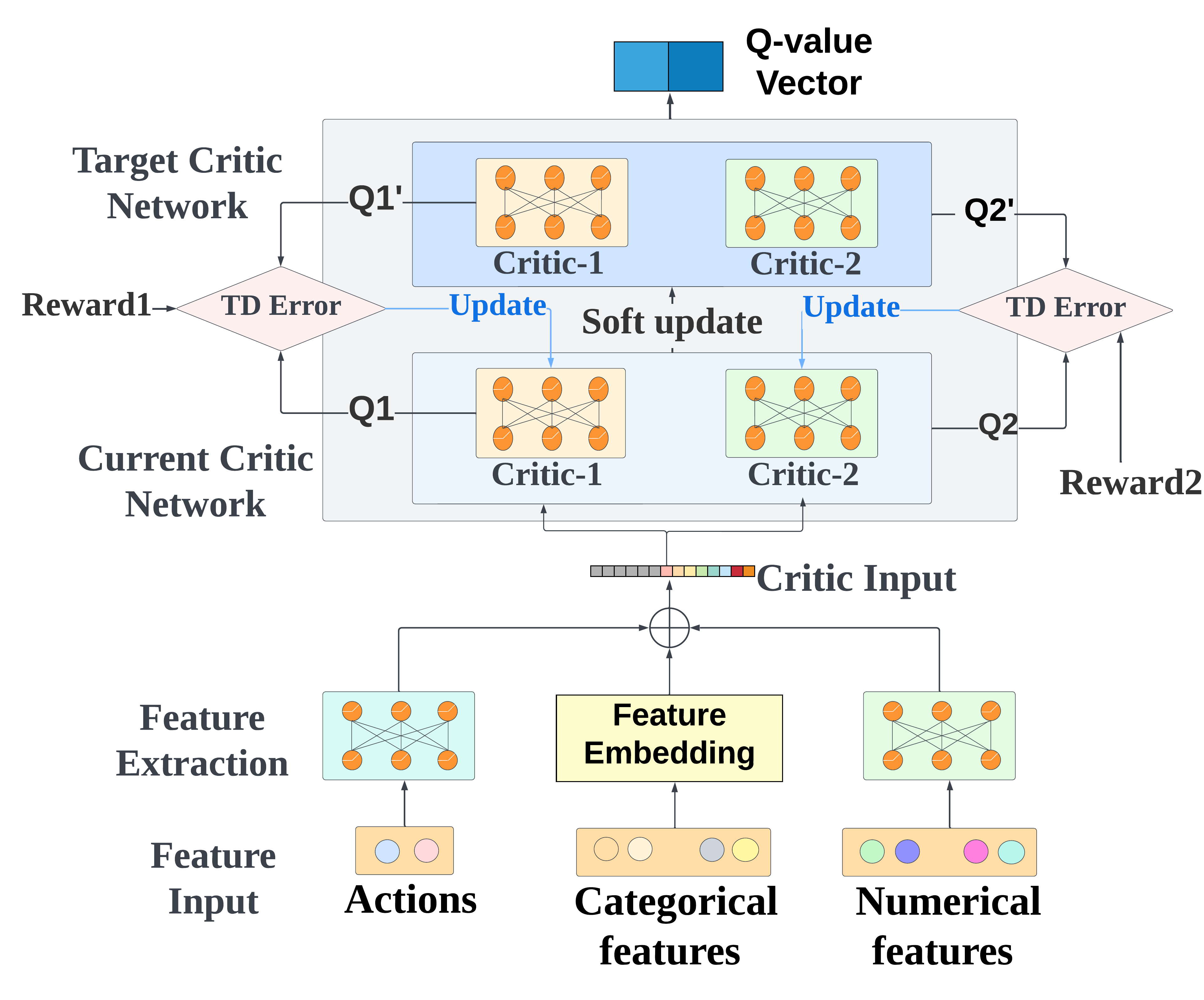}
    \vspace{-8mm}
    \caption{Structure of Critic Network.}
    \label{fig:Critic}
    \vspace{-5mm}
\end{figure}

The traditional critic network estimates the action value $Q$ generated from actor network and then updates the actor network parameters based on value $Q$. While we choose the MTL model as the actor network, the problem is how to design a suitable critic network structure for multi-parameter updating. In this paper, we propose a multi-critic structure with two parallel MLP networks sharing one bottom layer. This design is capable of updating MTL model parameters in the actor network and polishing the loss function weights for specific tasks. The structure of the critic network is shown in Figure \ref{fig:Critic}.
The first part of the critic network is one shared-bottom layer, which transforms the user-item features and action information simultaneously. Similar to the state representation network in Sub-subsection \ref{sss:srn}, we apply one embedding layer and an MLP structure for feature extraction. 
We then combine the user-item feature and action information as the input of two differentiable action value networks parameterized by $\phi_k$, which output the estimated Q-value based on the state-action information for each task. Given the current state $s_t$ and action $a_{k,t}$, the Q-value is calculated by:
\begin{eqnarray} \label{Q}
Q'(s_t,a_{k,t})
&=& \mathop{\mathbb{E}}[r_{k,t}+\gamma V(s_{t+1})|s_t,a_{k,t}] \\
&=& r_{k,t}+\gamma \sum_{s^{t+1} \in \mathcal{S}} p_{s_t,a_t, s_{t+1}} \cdot \max Q'(s_{t+1},a_{k,t+1})  \nonumber 
\end{eqnarray}
where $V(s_{t+1})$ is state value function. In our case, the action value $a_{t,k}$ for task $k$ is estimated by the actor network, and the next state $s_{t+1}$ is determined with a probability equal to 1. Therefore, the Q-value function in multi critic structure can be calculated as:
\begin{eqnarray} \label{Q}
Q(s_t,a_{k,t};\phi_k)
&=& r_{k,t}+\gamma Q(s_{t+1},a_{k,t+1};\phi_k) 
\end{eqnarray}
The weight of the objective loss function in Equation~\eqref{eq:swl} is reversely adjusted along the direction of the Q value to improve the optimization process of actor network, which is the linear transformation with punish variable $\lambda$:
\begin{equation}
    \omega_{k,t} = 1-\lambda*Q(s_t,a_{k,t}; \mathbf{\phi}_k)
\end{equation}
where $a_{k,t}$ is an action for task $k$ at timestamp $t$. $Q(s_t,a_{k,t}; \mathbf{\phi}_k)$ is action value of state $s_t$ and action $a_{k,t}$.

\subsection{Optimization}
The general framework of the actor-critic network usually encounters an inevitable problem: the convergence of the critic network is not guaranteed. To overcome this problem, we leverage the idea of Deterministic Policy Gradient Algorithms ~\cite{silver2014deterministic}, introducing the \textit{target network} into the learning framework. The target networks have exactly the same structure as the proposed actor and critic networks (i.e., \textit{estimation networks}), and we denote them as $\pi(s_t; \widetilde{\theta_k})$ and $Q(s_t,a_{k,t}; \mathbf{\widetilde{\phi}_k})$, which have lagging parameters $\widetilde{\theta}_k$ and $\widetilde{\phi}_k$. In the rest part of this paper, we specify the main actor-critic network by estimation actor networks parameterized by $\theta_k$ and estimation critic networks parameterized by $\phi_k$. The traditional actor-critic network usually suffers from the problem of parameter divergence and the convergence of the critic network is not guaranteed. The target network can address these problems by generating stationary target values from logging parameterized neuron networks, which smooths the training process. Next, we detail the optimization process for model parameters of the estimation actor-critic network, and also the soft update for target networks.


\noindent \textbf{Estimation critic network updates.} $\phi_k$ is the crucial parameter in the critic network, which determines action Q-value. Given transition $(s_t,a_{1,t},a_{2,t},s_{t+1},r_{1,t},r_{2,t})$ 
from replay buffer $\mathcal{B}$, the TD target of the target critic network for $k$-th task is derived from:
    \begin{equation}
        TD_{k,t} = r_{k,t} + \gamma Q(s_{t+1},a_{k,t+1}; \widetilde{\phi}_k)
    \end{equation}
    where $a_{k,t+1} = \pi(s_{t+1};\mathbf{\widetilde{\theta}}_k)$ is the estimated next action from target actor network.
    The Q-value generated from the estimation critic network, which estimates the current action value is defined as:
    \begin{equation}
        Q_{k,t} = Q(s_{t},a_{k,t}; \mathbf{\phi}_k)
    \end{equation}
    
\noindent After the training process among $b$ batch of transitions from the replay buffer, we calculate the average TD error $\delta$:
\begin{small}
    \begin{eqnarray} \label{eq}
        \delta 
        &=& \frac{1}{2|b|} \sum\limits_{(s_t,\cdots,s_{t+1},\cdots)\in b} \sum\limits_{k=1}^2 (TD_{k,t}-Q_{k,t}) \\
        &=& \frac{1}{2|b|} \sum\limits_{(s_t,\cdots,s_{t+1},\cdots)\in b} \sum\limits_{k=1}^2 \nonumber \\
        &\quad& [r_{k,t} + \gamma Q(s_{t+1},a_{k,t+1}; \widetilde{\mathbf{\phi}}_k)-Q(s_{t},a_{k,t}; \mathbf{\phi}_k)] \nonumber 
    \end{eqnarray} 
\end{small}
Then we update the current critic network for each task by the following gradient decent method:
    \begin{equation}
        \phi_{k}^{t+1} = \phi_{k}^{t} - \alpha^{\mathbf{\phi}}\mathbf{I}\delta \nabla_{\mathbf{\phi_k}}Q(s_{t},a_{k,t}; \mathbf{\phi}_k)
    \end{equation}
    where $\nabla_{\mathbf{\phi_k}}Q(s_{t},a_{k,t}; \mathbf{\phi}_k)$ is the gradient of $k$-th target Q-value. This completes the optimization of the estimation critic networks.

\noindent \textbf{Estimation actor network updates.} Before the TD error $\delta$ converges to threshold $\epsilon$, we update the current actor networks parameterized by $\theta_k$ through the gradients back-propagation of loss function for each tower layer after the forward process of each batch transitions from $\mathcal{B}$:
    \begin{equation}
        \theta_{k}^{t+1} = \theta_{k}^{t} + \alpha^{\mathbf{\theta}}\mathbf{I} \nabla_{\mathbf{\theta}_k} \mathcal{J}(\theta_k)
    \end{equation}
    where the loss for tower layers is defined by the negative of average Q-value $\mathcal{J}(\theta_k) = - \frac{1}{b}\sum\limits_{(s_t,\cdots,s_{t+1},\cdots)\in b} Q'(s_t,\pi(s_t; \mathbf{\theta}_k))$.
    \noindent We further update the current actor networks with respect to the gradients of the overall loss function of CTR/CTCVR prediction:
    \begin{equation}
        \theta_{k}^{t+1} = \theta_{k}^{t} - \alpha^{\mathbf{\theta}}\mathbf{I} \nabla_{\mathbf{\theta}_k} \mathcal{L}(\theta_1,\theta_2)
    \end{equation}
    \begin{equation}
        \mathcal{L}(\theta_1,\theta_2) = \sum\limits_{(s_t,\cdots,s_{t+1},\cdots)\in b} \sum\limits_{k=1}^2 \omega_{k,t} BCE(\pi(s_t; \mathbf{\theta}_k),y_{k,t})
    \end{equation}
where the weights $\omega_{k,t} = 1-\lambda*Q(s_t,a_{k,t}; \mathbf{\phi}_k)$ controlled by the corresponding current critic network, which is  adaptively adjusted to the negative direction of estimated action Q-value at $t$.

\noindent \textbf{Soft update of the target network.} 
    The target actor network and two specific target critic networks are updated until the current critic network reaches the convergence condition towards the direction of parameters in current networks:
    \begin{equation}
    \begin{split}
        \widetilde{\theta}_k = \beta \widetilde{\theta}_k + (1-\beta)\theta_k \\
        \widetilde{\phi}_k = \beta \widetilde{\phi}_k + (1-\beta)\phi_k
    \end{split}
    \end{equation}
    where $\beta \in [0,1]$ is the soft update rate.

\section{Experiment}
In this section, we conduct several experiments using two real-world datasets to evaluate the effectiveness of the RMTL framework. 

\subsection{Experimental Setup}

\subsubsection{\textbf{Dataset}.}
From our survey, there is only one open-source dataset, RetailRocket\footnote{https://www.kaggle.com/datasets/retailrocket/ecommerce-dataset}, that contains sequential labels of click and pay. 
In order to have a comprehensive comparison of RMTL, we also apply the ``Kuairand-1K'' dataset\footnote{https://kuairand.com/} \cite{gao2022kuairand} for further analysis. Each dataset is split into training, validation, and testing sets with proportion 6:2:2 sorted by timestamp $t$ for model learning.


\subsubsection{\textbf{Baselines.}}
Since our RMTL structure modifies the MTL objective loss function, we choose the MTL models with their default losses and one RL-based model as our baselines.
\textbf{Single Task} \cite{luong2015multi}: The naive method to learn each task separately, which is widely used to compare the performance of MTL models. 
\textbf{Shared Bottom} \cite{ma2018mmoe}: The basic structure of MTL models with a shared bottom layer.
\textbf{ESMM} \cite{ma2018esmm}: A model specified in CTR/CVR predictions that focus on the different sample spaces for each task.
\textbf{MMoE} \cite{ma2018mmoe}: MMoE uses gates to control the relationship between the shared bottom and the task-specific towers.
\textbf{PLE} \cite{tang2020ple}: PLE is able to capture complicated task correlations by using shared experts as well as task-specified expert layers.
\textbf{D-PLE} \cite{silver2014deterministic}: We apply DDPG (deep deterministic policy gradient) with the PLE model as the RL-based baseline.
Actually, all the loss functions for the baseline models are obtained by taking the unchangeable weighted average.

\begin{table*}[]
        \small
        \Description{Performance on CTR/CTCVR tasks for different methods.}
	\caption{Performance on CTR/CTCVR tasks for different methods.}
	\vspace{-3mm}
	\label{table:result1}
\resizebox{\textwidth}{!}{%
	\begin{tabular}{@{}|c|c|c|ccccccccc|}
		\toprule[1pt]
		\multirow{3}{*}{Dataset} & \multirow{3}{*}{Task} & \multirow{3}{*}{Metric} & \multicolumn{9}{c|}{Methods}\\ 
  
            \cmidrule(l){4-12} 
	&  &  & Single & Shared & ESMM & MMoE & PLE & D-PLE  & RMTL- & RMTL- & RMTL- \\

            &  &  & Task & Bottom &  &  &  &  & ESMM & MMoE & PLE \\ \midrule

            \multirow{3}{*}{RetailRocket} & \multirow{3}{*}{CTR} 
		& AUC $\uparrow$ & 0.7273 & 0.7287 & 0.7282 & \underline{0.7309} & 0.7308 & 0.7308 & 0.7338* & \textbf{0.7350*} & 0.7339* \\
		&  & Logloss $\downarrow$ & 0.2065 & 0.2048 & 0.2031 & \underline{0.2021} & 0.2056 & 0.2058 & 0.2024 & \textbf{0.1995} & 0.2013 \\
            &  & s-Logloss $\downarrow$ & 0.0846  & 0.0839  & 0.0852  & 0.0853  & \underline{0.0827} & 0.0830 & 0.0836  & 0.0848  & \textbf{0.0824}  \\ \midrule

            \multirow{3}{*}{RetailRocket} & \multirow{3}{*}{CTCVR} 
		& AUC $\uparrow$     & 0.7250 & 0.7304 & 0.7316 & 0.7347 & \underline{0.7387} & 0.7386 & 0.7341 & 0.7396 & \textbf{0.7419*} \\
		&  & Logloss $\downarrow$ & 0.0489 & 0.0493 & 0.0486 & 0.0496 & \underline{0.0486} & 0.0490 & 0.0485 & 0.0490 & \textbf{0.0480} \\ 
            &  & s-Logloss $\downarrow$ & 0.0150  & 0.0149  & 0.0150  & \underline{0.0145}  & 0.0147 & 0.0149  & 0.0150  & \textbf{0.0143}  & 0.0146  \\ \midrule

            \multirow{3}{*}{Kuairand} & \multirow{3}{*}{CTR} 
		& AUC $\uparrow$          & 0.7003  & 0.7018  & 0.7009  & 0.7014  & \underline{0.7026} & 0.7025 &  0.7031  & 0.7029  & \textbf{0.7053*}  \\
		&  & Logloss $\downarrow$ & 0.6127  & 0.6114  & 0.6128  & 0.6119 & \underline{0.6111} & 0.6123 & 0.6111  & 0.6105  & \textbf{0.6092*}  \\ 
            &  & s-Logloss $\downarrow$ & 0.6263 & \underline{0.6250}  & 0.6261  & 0.6255  & 0.6252 & 0.6257 & 0.6252 & 0.6243  & \textbf{0.6242}  \\ \midrule

            \multirow{3}{*}{Kuairand} & \multirow{3}{*}{CTCVR} 
		& AUC $\uparrow$          & 0.7342 & 0.7310  & \underline{0.7350}  & 0.7324  & 0.7339 & 0.7339 & \textbf{0.7377*}  & 0.7345  & 0.7367*  \\
		&  & Logloss $\downarrow$ & 0.5233 & 0.5249  & \underline{0.5220}  & 0.5237  & 0.5235 & 0.5237  & \textbf{0.5200*}  & 0.5225  & 0.5221  \\ 
            &  & s-Logloss $\downarrow$ & 0.5449  & 0.5468  & \underline{0.5436}  & 0.5450  & 0.5454 & 0.5451  & \textbf{0.5433}  & 0.5446  & 0.5447  \\ \midrule
	\end{tabular}}
	\\ $\uparrow$: higher is better; $\downarrow$: lower is better. \underline{Underline}: the best baseline model. \textbf{Bold}: the best performance among all models. 
	\\``\textbf{{\Large *}}'': the statistically significant improvements (i.e., two-sided t-test with $p<0.05$) over the best baseline.
	\vspace{-3mm}
\end{table*}

\subsubsection{\textbf{Evaluation Metrics.}}
\begin{itemize}[leftmargin=*]
    \item \textbf{AUC} and \textbf{Logloss}: These are the straightforward metrics to evaluate CTR/CTCVR prediction tasks. According to \cite{guo2017deepfm}, it is statistically significant in recommendation tasks that AUC or Logloss changes more than 0.001-level.
    \item \textbf{s-Logloss}: This metric is defined as the averaged Logloss with respect to all sessions.
\end{itemize}


\vspace{-2mm}
\subsection{Implementation Details}
We implement the baseline models using the public library\footnote{https://github.com/easezyc/Multitask-Recommendation-Library.git}, which integrates 7 standard MTL models (without ESMM) and gives a reference performance of each model. All the models have the same basic network structure, i.e., an input embedding layer with dimension 128, a $128 \times 512 \times 256$ Multi-Layer Perceptron (MLP) as the bottom layer, and a $256 \times 128 \times 64 \times 1$ MLP as the tower layer. Specifically, for the models using expert layers (same structure as the bottom layer), we fix the number of experts as 8. The activation function for the hidden layers is ReLU and Sigmoid for the output. We also set a $0.2$ dropout rate. We use Adam optimizer ($lr = 1e-3$) to learn all the models. We also implement ESMM sharing the hyperparameters of the above models.

Since the actor network in our RMTL model is a two-tower design with multi-task output, it is compatible with most existing MTL-based recommendation models. In this paper, we apply our RMTL structure on three representative MTL models: ESMM, MMoE, and PLE.
As to the RMTL structure, we set the Actor network with the same structure as specified MTL models. The Critic networks also share the same bottom layer and have their out through one tower layer. Moreover, we add a $1 \times 128$ action layer to be consistent with the input features, and we change the output activation to negative ReLU in that our reward only has negative values. In order to improve data efficiency, we initialize the Actor with pretrained parameters from the MTL models and keep them frozen until the Critics converge. Then we multiply the critic value in the total loss and retrain the MTL model. The default learning rates of the actor $\alpha^{\theta}$ and critic network $\alpha^{\phi}$ are set as 0.001. We set the default soft update rate $\beta = 0.2$, punish variable $\lambda = 0.7$. 
The implementation code is available online to ease reproducibility\footnote{https://github.com/Applied-Machine-Learning-Lab/RMTL}.

\subsection{Overall Performance and Comparison}
We compare the performance of five baseline MTL models with RMTL models for CTR/CTCVR prediction tasks on two different datasets. The overall performance on CTR/CTCVR tasks for different methods is shown in Table~\ref{table:result1}. It can be observed that:
\begin{itemize}[leftmargin=*]
\item The SingleTask model achieves the worst performance in both prediction tasks on two datasets. The feature representation network and loss function optimization are separated for each task, which fails to detect the intrinsic relation between multiple tasks. Thus, we can understand that the shared-bottom method, which shares the feature extraction parameters, can outperform the SingleTask method for both datasets. 
\item The PLE model achieves almost the best performance among all MTL baseline models in most cases. On the basis of the MMoE model, which controls the parameter interaction between the shared bottom and specific task tower, the PLE model specifies the parameter sharing between expert layers to extract hidden interaction patterns between each task. This demonstrates that the PLE baseline model can improve the efficiency of information sharing among tasks to achieve better prediction performance. 
\item Each version of our proposed RMTL model outperforms the corresponding non-RL version baseline model (e.g., RMTL-ESMM v.s. ESMM) in both prediction tasks with AUC and logloss metrics on two datasets. Especially on the RetialRocket dataset, the RMTL model achieves around 0.003-0.005 AUC gains compared with the corresponding baseline model. By leveraging the sequential property of the reinforcement learning framework, the RL-enhanced method is capable of processing session-wise recommendation data and achieves significant improvement in CTR/CTCVR prediction tasks by adaptively adjusted loss function weights. 
\item The CTCVR prediction result on the Kuairand dataset presented in the fourth row of Table \ref{table:result1}, where the ESMM baseline model and its advanced RL version achieve better performance than other MTL models. For short video recommendations, click-and-convert behaviors usually have a relatively higher correlation. In this phenomenon, the ESMM model outperforms MMoE and PLE by overcoming the sample selection bias.
\item The RMTL models achieve s-Logloss improvement of less than 0.001 compared with the corresponding baseline models, which is lower than that of Logloss. This phenomenon may be caused by the fact that session-based metric has similar performance with online A/B test, which is more difficult to be improved \cite{DBLP:journals/corr/abs-2003-11941}. Note that the ratio of logloss and s-logloss is different for the two datasets since they have different average session lengths while the average session Logloss is affected by the average session length for a specific dataset.
\end{itemize}
To summarize, the RMTL model outperforms the state-of-the-art MTL models on both CTR and CTCVR prediction tasks for different real-world datasets. In addition, it can be employed in most MTL models, validating its desirable compatibility.

\begin{figure}[t]
    \Description{Transferability study results for ESMM, MMOE, and PLE on the RetailRocket dataset.}
    \centering
	{\subfigure{\includegraphics[width=0.327\linewidth]{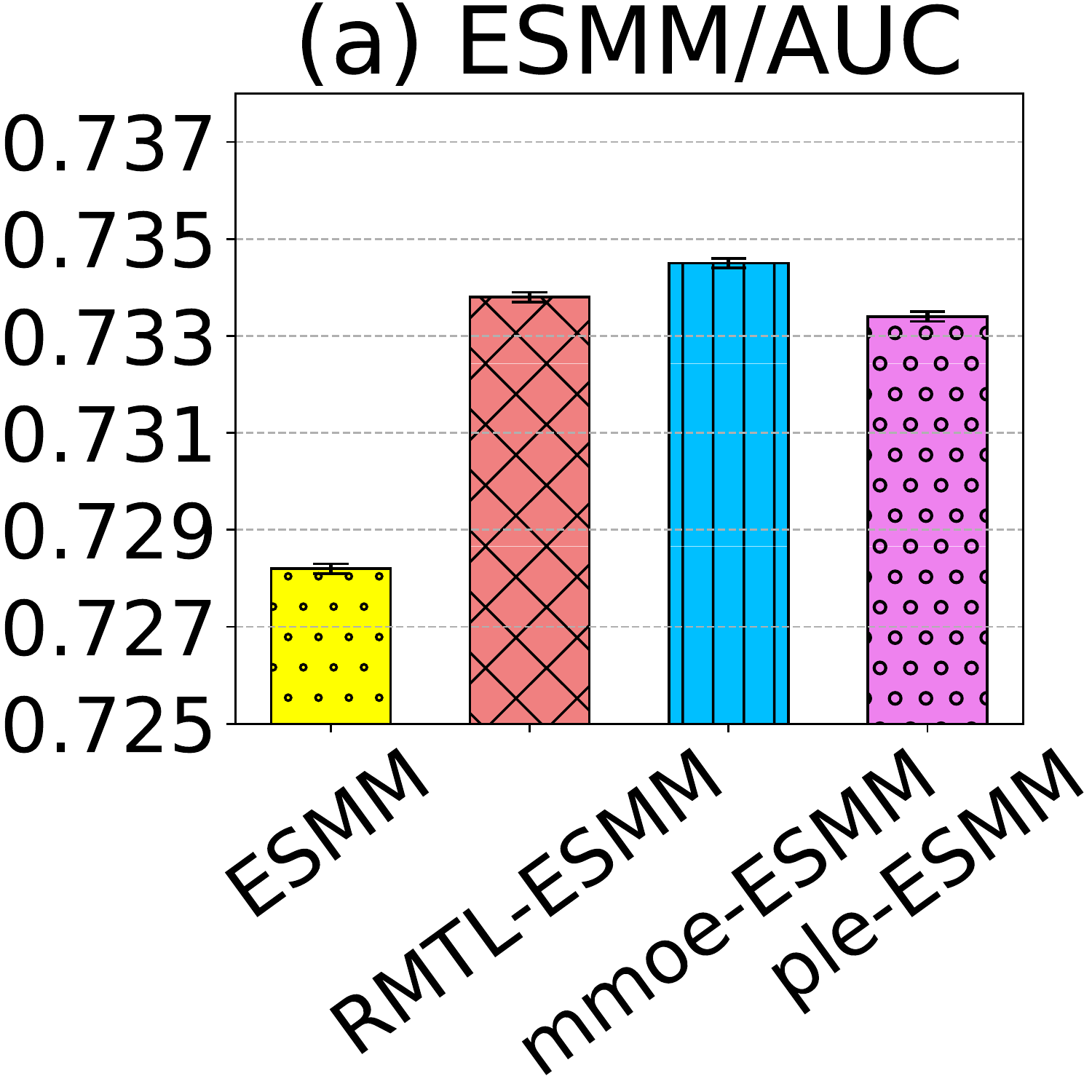}}}
	{\subfigure{\includegraphics[width=0.327\linewidth]{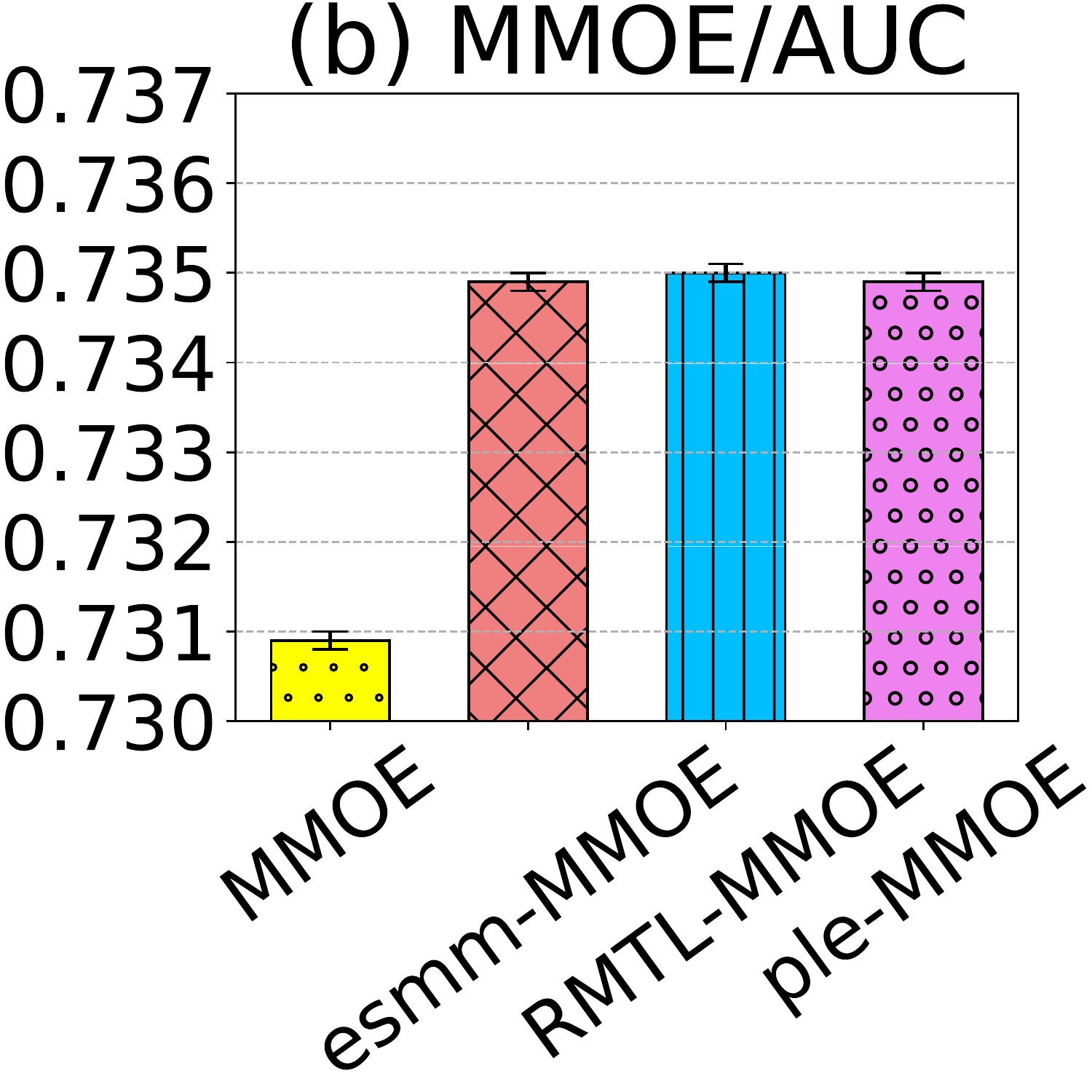}}}
	{\subfigure{\includegraphics[width=0.327\linewidth]{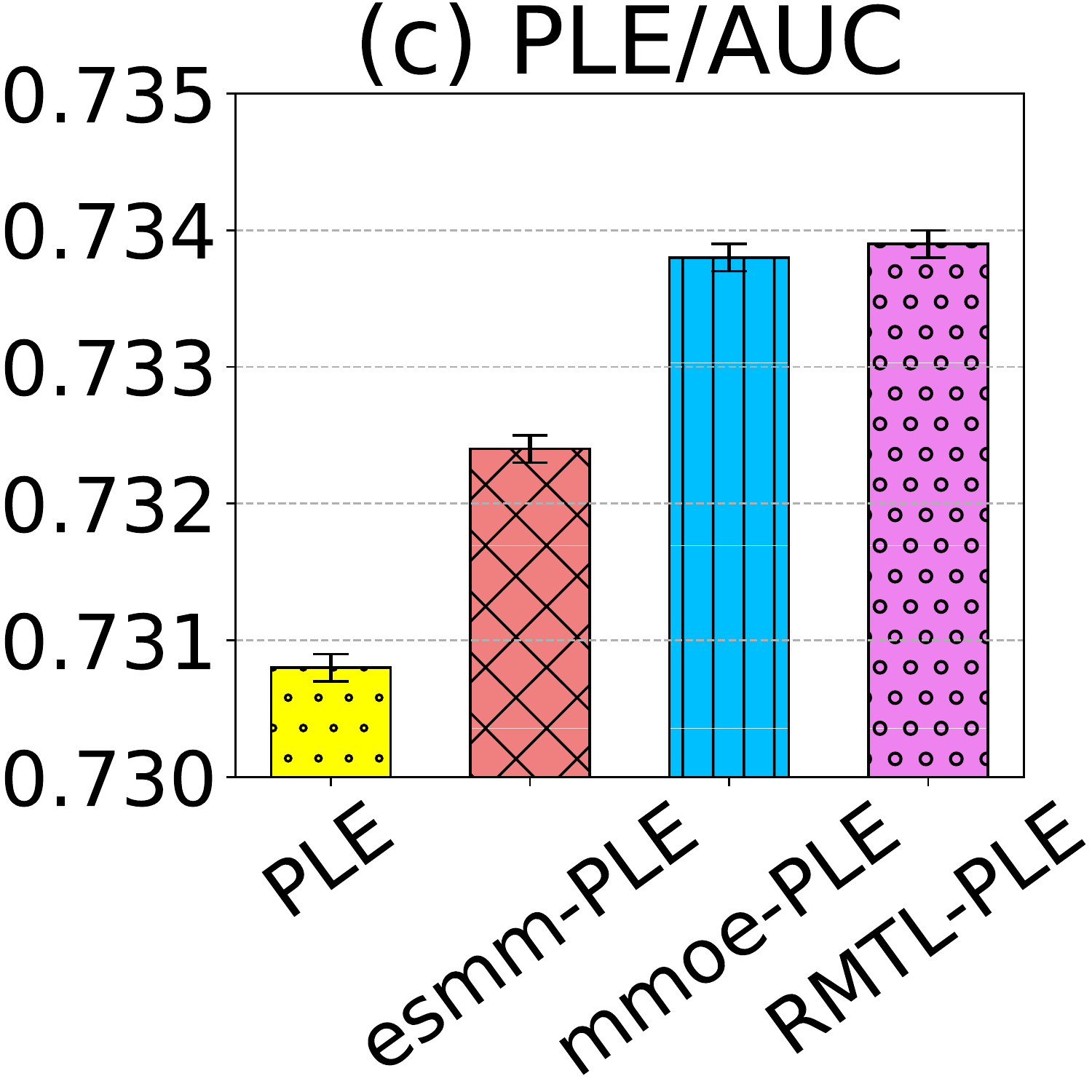}}}
	\vspace*{-3mm}
	{\subfigure{\includegraphics[width=0.327\linewidth]{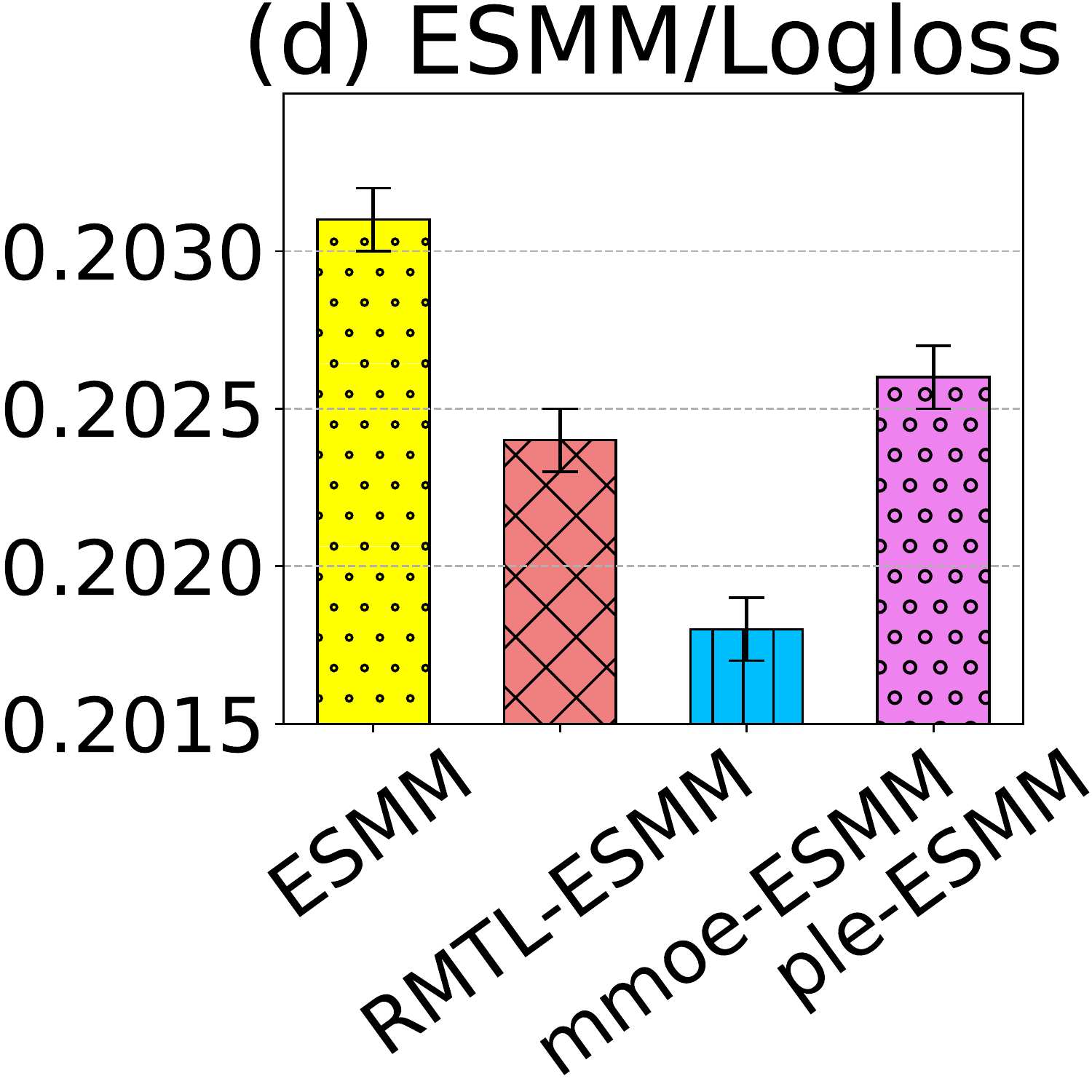}}}
	{\subfigure{\includegraphics[width=0.327\linewidth]{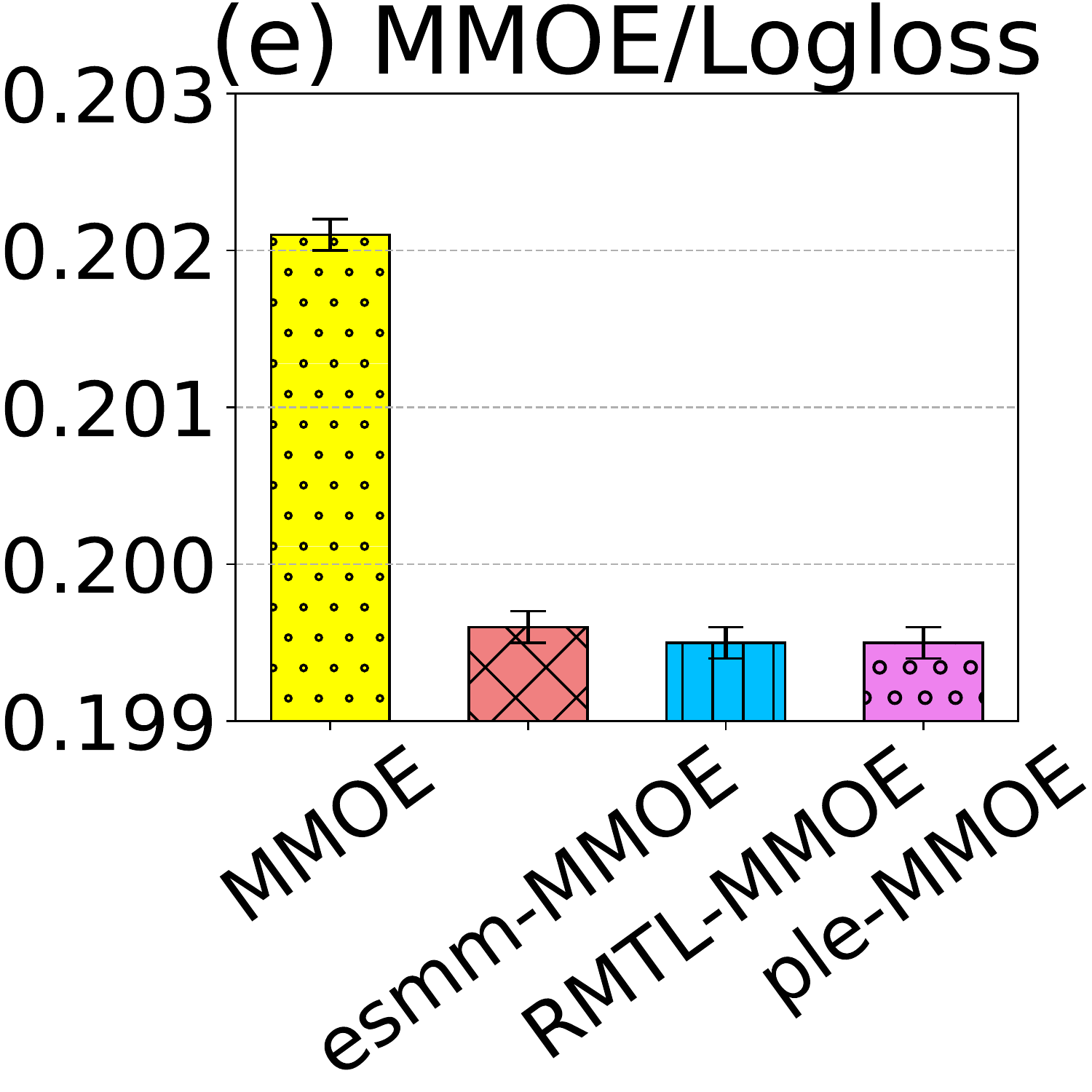}}}
	{\subfigure{\includegraphics[width=0.327\linewidth]{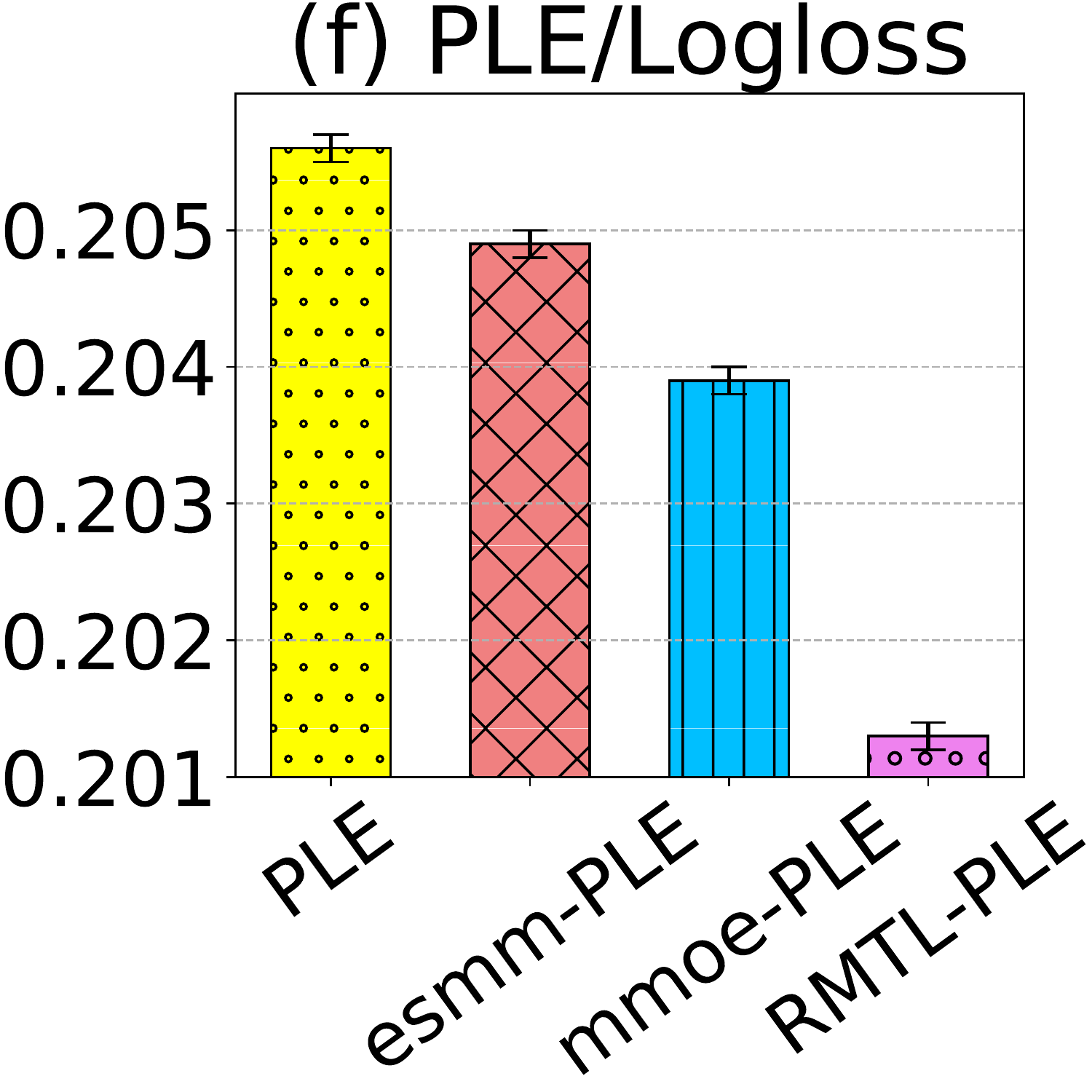}}}
	\vspace{-5.6mm}
	\caption{Transferability study results.}\label{fig:Fig3}
	\vspace{-4.9mm}
\end{figure}
\subsection{Transferability Study}
This subsection presents the transferability study for the RMTL method for the CTR task on the RetailRocket dataset. We try to figure out whether the critic network learned from different logging policies can be applied to the same MTL baseline model and improve the prediction performance. For each MTL baseline model, we leverage the model parameters from the critic network pretrained by ESMM, MMoE, and PLE model on the RetailRocket dataset. The results of AUC and logloss are presented in Figure \ref{fig:Fig3} (a)-(c) and (d)-(f), where ``mmoe-ESMM'' means ESMM model applying critic network trained from MMoE and ``ple-ESMM'' means ESMM model applying critic network trained from PLE. It can be observed that:
(i) The pretrained critic network from three MTL models can significantly increase AUC for each baseline model. 
(ii) The pretrained critic network from three MTL models can significantly decrease logloss for each baseline model.

To summarize, the pretrained critic network is capable of improving the prediction performance of most MTL models, which demonstrates the excellent transferability of the RMTL model.

\begin{table}[t]
        \Description{Ablation study on RetailRocket.}
	\caption{Ablation study on RetailRocket.}
	\vspace{-3mm}
	\label{table:result2}
	\begin{tabular}{@{}|c|c|cccc|@{}}
		\toprule[1pt]
		\multirow{3}{*}{Task} & \multirow{3}{*}{Metric} & \multicolumn{4}{c|}{Methods}\\ 
            \cmidrule(l){3-6} 
		&  & CW & WL & NLC & RMTL-PLE \\ \midrule

            \multirow{2}{*}{CTR} 
		& AUC $\uparrow$ & 0.7308 & 0.7323 & 0.7325 & \textbf{0.7339*} \\
		& Logloss $\downarrow$ & 0.2056 & 0.2218  & 0.2020 & \textbf{0.2013 *} \\ \midrule

            \multirow{2}{*}{CTCVR} 
		& AUC $\uparrow$   & 0.7387  & 0.7411  & 0.7418  & \textbf{0.7419*} \\
		& Logloss $\downarrow$ & 0.0486 & 0.0485 & 0.0481 & \textbf{0.0480*} \\ \bottomrule[1pt]
	\end{tabular}
	\\ ``\textbf{{\Large *}}'' indicates the best performance among all variants.
		\vspace{-5mm}
\end{table}

\subsection{Ablation Study}
The most essential design of the RMTL model is the  overall loss function with learnable weights, which is the linear combination of Q-value from the critic network. In this subsection, in order to figure out the importance of this loss function in the proposed model, we change some components and define three variants below:
(i) \textbf{CW:} The CW variant indicates applying constant weights for the  overall loss function and no gradient policy update for the actor network, which eliminates the contribution of the critic network. We assign all weights equal to constant 1.
(ii) \textbf{WL:} The WL variant indicates the adjustment of weights $\omega$ is controlled by session behavior labels $y_k$ multiplied by Q-value. 
(iii) \textbf{NLC:} No linear transformation is performed to the loss weights, instead, we directly assign the negative Q-value to loss weights.

The ablation study results for the PLE model on the RetailRocket dataset are shown in Table \ref{table:result2}. It can be observed that:
(i) CW has the worst performance for AUC and logloss metrics on both prediction tasks. It may be caused by equivalently assigned BCE loss weights, which fail to capture the hidden dynamics of multi-tasks.
(ii) WL and NLC have almost the same performance in this study that outperform the CW variant with 0.002-0.003 AUC gain. This demonstrates the effect of weights updating from the critic network.
(iii) RMTL-PLE with our proposed total loss setting achieves the best performance on both tasks, which illustrates the validity of our linear combination weight design.
We may summarize that our proposed MTL model can achieve better CTR/CTCVR prediction performance, where total loss design has great contributions.

\section{Related Work}\label{sec: related_work}
In this section, we give a brief introduction to existing works related to RMTL, i.e., RL-based and MTL-based recommender systems.

\noindent\textbf{Reinforcement Learning Based Recommender Systems.}
Plenty of research has been done combining Reinforcement Learning (RL) ~\cite{sutton2018reinforcement,afsar2021reinforcement,wang2022surrogate,zhang2022multi, zhao2017deep, zhao2018deep, zhao2021dear} into the field of Recommender Systems (RS).
Compared to traditional learning-to-rank solutions of RS ~\cite{liu2009learning} that optimize the immediate user feedback, the RL-based RS focuses on optimizing the cumulative reward function that estimates multiple rounds of interactions between the recommendation policy and the user response environment. 
The general problem formulates the sequential user-system interactions as a Markov Decision Process (MDP) ~\cite{shani2005mdp}.
RL solutions have been studied under this formulation including early-age tabular-based methods ~\cite{joachims1997webwatcher,mahmood2007learning,moling2012optimal} that optimize an evaluation table for a set of state-action pairs, value-based methods ~\cite{taghipour2007usage,zheng2018drn,zhao2018recommendations,ie2019slateq} that learn to evaluate the quality of an action or a state, policy gradient methods ~\cite{sun2018conversational,chen2019top,chen2019large} that optimize the recommendation policy based on the long-term reward, and the actor-critic methods ~\cite{liu2018deep} based on policy gradient method ~\cite{sutton1999policy,peters2008natural,bhatnagar2007incremental,degris2012model} that simultaneously learn an action evaluator and an action generator.
Our method falls into the actor-critic category and extends it toward the multi-task problem setting.
The main challenges of applying RL for the recommendation task consist of the large combinatorial state/actions space ~\cite{dulac2015deep,liu2020state}, the uncertainty of user environment ~\cite{ie2019recsim,zhao2019deep}, the stability and efficiency of exploration ~\cite{bai2019model,chen2021user}, and the design of a user's reward function over heterogeneous behaviors ~\cite{zou2019reinforcement}.
Our work is related to the reward function design but also enhances the performance of each task.

\noindent\textbf{Multi-Task Learning in Recommender Systems.}
As stated in ~\cite{zhang2021survey}, Multi-Task Learning (MTL) is a machine learning framework that learns a task-invariant representation of an input data in a bottom network, while each individual task is solved in one’s respective task-specific network and boosted by the knowledge transfer across tasks.
Recently, MTL has received increasing interest in recommender systems ~\cite{ma2018esmm,lu2018coevolutionary,hadash2018rank,pan2019mandarin,pei2019value} due to its ability to share knowledge among different tasks especially its ability to capture heterogeneous user behaviors. 
A series of works seek to improve on it by designing different types of shared layer architectures. 
These works either introduce constraints on task-specific parameters ~\cite{duong2015low,misra2016cross,yang2016deep} or separate the shared and the task-specific parameters ~\cite{ma2018mmoe, tang2020ple}.
The general idea is to disentangle and share knowledge through the representation of the input feature. 
Additionally, there is also research on applying multi-agent RL for the multi-scenario setting \cite{feng2018learning} where the recommendation task is bundled with other tasks like search, and target advertising.
Different from the above ideas, we resort to knowledge distillation to transfer ranking knowledge across tasks on task-specific networks and we combine RL to improve the long-term satisfaction of users. 
Notably, our model is a general framework and could be leveraged as an extension for most off-the-shelf MTL models. 

\section{Conclusion}
In this paper, we propose a novel multi-task learning framework, RMTL, to improve the prediction performance of multi-tasks by generating dynamic total loss weights in an RL manner. The RMTL model can adaptively modify the weights of BCE for each prediction task by Q-value output from the critic network. By constructing a session-wise MDP environment, we estimate the multi-actor-critic networks using a specific MTL agent and then polish the optimization of the MTL overall loss function using dynamic weight, which is the linear transformation of the critic network output. 
We conduct several experiments on two real-world commercial datasets to verify the effectiveness of our proposed method with five baseline MTL-based recommendation models. The results demonstrate that RMTL is compatible with most existing MTL-based recommendation models and can improve multi-task prediction performance with excellent transferability.

\section*{ACKNOWLEDGEMENTS}
This research was partially supported by APRC - CityU New Research Initiatives (No.9610565, Start-up Grant for New Faculty of City University of Hong Kong), SIRG - CityU Strategic Interdisciplinary Research Grant (No.7020046, No.7020074), HKIDS Early Career Research Grant (No.9360163), Huawei Innovation Research Program and Ant Group (CCF-Ant Research Fund).

\bibliographystyle{ACM-Reference-Format}
\bibliography{ref}


\end{document}